\documentclass[nofootinbib]{revtex4}  
\usepackage{hyperref}
\usepackage{amssymb,amsmath,amsfonts,latexsym,amsbsy}
\renewcommand{\vec}[1]{\ensuremath{{\boldsymbol{#1}}}}

\newcommand{\R}{{\vec{\rm R}}}
\newcommand{\xv}{{\vec{\rm x}}}
\newcommand{\ka}{{\vec{\rm k}}}

\newcommand{\q}{{\vec{\rm q}}}
\newcommand{\p}{{\vec{\rm p}}}
\newcommand{\vs}{{\vec{\rm v}}}
\newcommand{\n}{{\vec{\rm n}}}

\newcommand{\lv}{{\vec{\rm l}}}

\newcommand{\jv}{\vec{\rm J}}
\newcommand{\kap}{\vec{\kappa}}

\newcommand%
{\mVV}%
[1]%
{{\langle #1 \rangle}}%
\newcommand%
{\MVV}%
[1]%
{{\langle\!\langle #1\rangle\!\rangle}}%
\newcommand%
{\MMV}%
[1]%
{{\left\langle\!\!\!\left\langle #1\right\rangle\!\!\!\right\rangle}}%
\begin{document}

\title{  Corrections to Scattering Processes  due to Minimal Measurable Length}

\author{Mir Faizal${}^{c,d}$ }
\email{mirfaizalmir@googlemail.com} 
\author{S.~E.~Korenblit${}^{a,b}$}
\email{korenb@ic.isu.ru }
\author{A.~V.~Sinitskaya${}^{a,f}$ }
\email{lasalvia@mail.ru}
\author{Sudhaker Upadhyay${}^{e}$ }
\email{sudhakerupadhyay@gmail.com}
\affiliation {${}^{c}$Department of Physics and Astronomy, 
University of Lethbridge, Lethbridge, AB T1K 3M4, Canada}
\affiliation{${}^{d}$Irving K. Barber School of Arts and Sciences, University of British 
Columbia - Okanagan, 3333 University Way, Kelowna, British Columbia V1V 1V7, Canada}
\affiliation {${}^{a}$Department of Physics, Irkutsk State University, 20 Gagarin blvd, 
Irkutsk 664003, Russia} 
\affiliation {${}^{f}$Irkutsk National Research Technical University, 83 Lermontov str., 
Irkutsk 664074, Russia}
\affiliation {${}^{b}$Joint Institute for Nuclear Research, RU-141980, Dubna, Russia}
\affiliation{${}^{e}$Department of Physics, K.L.S. College, Nawada-805110, Bihar, India}
\begin{abstract} 
In this paper, we will analyze the short distance corrections to  low energy scattering. 
They are produced because of an intrinsic extended structure of the background geometry of 
spacetime. 
It will be observed that the deformation produced by a minimal measurable length can have low 
energy consequences, if this extended structure occurs at a scale much larger than the Planck 
scale.
We explicitly calculate short distance corrections to the Green function of the deformed 
Lippmann-Schwinger equation, and to the conserved currents for these processes. 
We then use them to analyze the pre-asymptotic corrections to the differential scattering flux 
at finite macroscopically small distances. 
\end{abstract}
\maketitle
We do not have a complete theory of quantum gravity, however, there are various approaches 
to quantum gravity.
It is  expected from various different approaches to quantum gravity that the geometry of spacetime 
could  be deformed by the existence of a minimal measurable length scale  \cite{s16}.
In fact, it is known that  in string theory, the background geometry of spacetime    gets 
deformed by the existence of a such minimal measurable length \cite{z2,zasaqsw}. 
The reason is that the smallest probe available in string theory is the fundamental string, 
and so the spacetime cannot be probed below the string length scale 
\cite{cscds,2z}. In fact,  it has been demonstrated that in perturbative string theory, the  minimal measurable length $l_{min}$
is related to the string length as  
 $l_{min} = g_s^{1/4} l_s$  (where $l_s = \alpha'$ is the string length, and 
 $g_s$ is the string coupling constant). Even though non-perturbative effects can produce 
 point like objects (such a $D0$-branes), it can be argued that  a  minimal length of the order of 
$l_{min} = l_s g_s^{1/3}$ is produced by these non-perturbative effects    \cite{s16, s18}. 
Such a minimal measurable length exists in string theory because the total energy of the quantized string depends on the winding number 
$w$ and the excitation number $n$. Now under T-duality, as $\rho\to l_s^2/\rho$, we have 
$n\to w$. 
Thus, it is possible to argue using the T-duality that a description of string theory 
below and above $l_s$  are the same, and so string theory  contains a minimal 
measurable length scale  \cite{s16}. It should be noted that an effective path integral of
the center of mass of the string (for strings propagating in compactified extra dimensions) 
has been constructed, and T-duality has been used to demonstrate that such a system has a 
minimal length associated with it \cite{green1, green2}. As the  construction of 
double field theory has been motivated from T-duality \cite{df12, df14}, 
it is expected that such a minimal length will also  exist in the double field theory  \cite{mi15}. 
  
It may be noted that even if the string theory does not turn to be the true theory of quantum 
gravity, the argument for the existence of a minimal measurable length in spacetime could 
still hold. As   it can be argued, a minimal measurable length scale, at least of the order 
of Planck length, would 
exist in all approaches to quantum gravity. This is because any theory of quantum gravity 
has to produce consistent black hole physics, and the black hole physics can be used to 
prove the existence of a minimal measurable length of the order of  Planck length. 
The reason is that the energy needed to probe spacetime below Planck length is less than 
the energy needed to produce a mini black hole region of spacetime \cite{z4,z5}.  
As production of such a mini black hole would restrict our ability to probe this region 
of spacetime, so the black hole physics can also predicts the existence of a minimal 
measurable length. 
In fact, it has been demonstrated that 
an extended structure in the background geometry of spacetime also exists in the loop quantum 
gravity \cite{z1}, and it is responsible for removing the big bang singularity. 
Furthermore, a minimal measurable length exists even in  Asymptotically Safe Gravity 
\cite{asgr}. It has also been argued that such a minimal measurable length will exist in 
conformally quantized quantum gravity \cite{cqqg}.  As it is possible than such a minimal measurable length exists in spacetime, 
 it is important to study the 
consequences of the existence of such a  minimal measurable length. 

As it is possible for the string length scale to be several orders larger than the Planck 
scale \cite{s16}, the minimal length scale can also  be much larger than the Planck length. 
In fact, it is possible to argue that in most models of quantum gravity such a minimal  
length can be much larger than Planck length (and it would only be bounded by the present 
experimental data) \cite{ml12, ml21}. 
It has been suggested that such a minimal measurable length much larger than Planck scale 
should have a measurable effects, which can be detected by performing more precise 
measurement of Landau levels and Lamb shift \cite{ml15}. 
Actually, it has also been proposed that such a minimal measurable length will deform 
quantum systems, and this deformation can be detected experimentally using 
an opto-mechanical setup \cite{ml14}. 
As this deformation can even be detected in precise measurement of low energy systems, it 
can also be detected in special future scattering experiments. So, it is important to 
consider the corrections to various scattering processes from such a minimal length. 
It has been suggested that the interaction between neutrons and a gravitational field can be measured using 
a gravitational spectrometer \cite{ne12ab, ne14ab}. 
The deformation of such a system by the minimal measurable length, and its possible detection using such 
a gravitational spectrometer, has also been studied  \cite{ml16}.
In fact, it has also been possible to obtain the  
corrections to quantum field 
theories and gauge theories (including standard model) from such a deformation 
\cite{ qft2, qft4, qft5, qft6, qft7}. Thus, it is important to analyze the 
effect this deformation will have on scattering processes. 
So, in this paper, we will analyze the modifications to a low energy scattering process by 
the existence of such an minimal measurable length  in spacetime. 
This minimal measurable length acts as an extended structure in spacetime, and 
the scattering of an extended structures is very different from the scattering of point 
particles. However, if the extended structure exists at a very small scale, then at large 
scale these phenomena can be expressed as a scattering of point particles. 
The corrections to these phenomena will occur at intermediate scales, and those will be 
the corrections we will analyze in this paper. This analysis implies an accurate 
description of internal finite (macroscopically small) distance corrections to the 
scattering process itself, that will be done here. 

It may be noted that it is possible for the deformation to occur at a scale much larger than 
Planck scale, and this scale would be bounded by the current experimental data 
\cite{ml12, ml21, ml15, ml14, ml16}. However, a deformation at such a scale would have 
measurable consequences for low energy phenomena, and this will hold for accurate measurements 
made on even non-relativistic quantum mechanical systems  \cite{ml12, ml21, ml15, ml14, ml16}.
So, we will now study such corrections to a non-relativistic scattering of a scalar particle 
by a Hermitian spherically symmetric potential $V(R)$. 
As we will be analyzing  a non-relativistic systems, so we will deform the system by a three 
dimensional spatial length   \cite{ml12, ml21, ml15, ml14, ml16}, and not a full four 
dimensional length in spacetime \cite{ qft2, qft4, qft5}. 
Even though we will consider only a single  scattering 
process, similar corrections will occur for any scattering 
process, as these corrections are induced by  an internal extended 
structure in spacetime. Thus, the  form of these corrections will be a universal feature of 
scattering processes, and  the main results of this paper can be used to obtain short
distance corrections to any scattering processes. 

To analyze the effect of the deformation on scattering processes, setting $\hbar=1$, 
we will take into account  the internal finite distance corrections, that are intrinsic 
to the scattering process itself. 
We note that for such simplest non-relativistic scattering on potential $V(R)$ the 
differential cross-section can be uniquely defined by on-shell scattering amplitude 
$f^+(\q;\ka)$ \cite{T,Ng}, when the initial momentum state $|\ka\rangle$ turn to the final 
momentum state $|\q\rangle$. The scattering amplitudes $f^\pm(\q;\ka)$ are coefficients of  
outgoing or incoming spherical waves as the first order terms of asymptotic expansion of the 
scattering wave function at the distance $R=|\R|\to\infty$:
\begin{eqnarray}
&&\!\!\!\!\!\!\!\!\!\!\!\!\!\!\!\!\!\!\!\!\!\!\!\! 
\Psi^\pm_\ka(\R)\underset{R\to\infty}{\longmapsto} 
e^{i(\ka\cdot\R)} + \frac{e^{\pm ikR}}{R}\,f^\pm(\q;\ka) +O(R^{-2}),\;\mbox { with }\;
\q=k\n,\;\; \ka=k\kap,\;\; \n^2=\kap^2=1,
\label{1_} 
\end{eqnarray}
where $\R=R\n$ for the spherical coordinates with spherical angles $\vartheta,\phi$ and  
$\n(\vartheta,\phi)=\vec{\rm e}_m{\rm n}_m \mapsto 
(\sin\vartheta\cos\phi,\sin\vartheta\sin\phi,\cos\vartheta)$ in Cartesian basis. 
The usual elastic differential and total cross-sections are defined by
\begin{eqnarray}
&&\!\!\!\!\!\!\!\!\!\!\!\!\!\!\!\!\!\!\!\!\!\!\!\! 
d\sigma =\left|f^+(\q;\ka)\right|^2 d\Omega(\n),\;\mbox {  and }\; 
\sigma =\int \left|f^+(k\n;k\kap)\right|^2 d\Omega(\n).
\label{3_}
\end{eqnarray}
These cross-sections do not depend on $R$, and this behavior is important in the quantum 
scattering theory \cite{T,Ng}.  
In order to discuss the full asymptotic expansion of scattering wave function, we note that  
$\Psi^\pm_\ka(\R)$ is a solution to Schr\"odinger equation for the scattering energy $E>0$ 
with $k^2=2ME$, potential $U(R)=2MV(R)$, and the vector operator 
$\vec{\nabla}_{\R}=\n\partial_R+R^{-1}\vec{\vec{\heartsuit}}_{\n}$ (with the usual angles 
dependent vector part $\vec{\vec{\heartsuit}}_{\n}\mapsto
(\partial_\vartheta,(\sin\vartheta)^{-1}\partial_\phi)$ in spherical basis \cite{blp}): 
\begin{eqnarray}
\left(\vec{\nabla}^2_{\R}+k^2\right)\Psi^\pm_\ka(\R)=U(R)\Psi^\pm_\ka(\R).
\label{4_}  
\end{eqnarray}
It satisfies also the respective Lippmann-Schwinger equation, with $\xv=\varrho\vs$, $\vs^2=1$      
\begin{eqnarray}
\Psi^\pm_\ka(\R)=e^{i(\ka\cdot\R)}-
\int\! d^3{\rm x}\,\frac{e^{\pm ik|\vec{\rm R}-\xv|}}{4\pi|\vec{\rm R}-\xv|}\,
U(|\xv|)\Psi^\pm_\ka(\xv). 
\label{5_}
\end{eqnarray}
Now with the operator of angular momentum 
$\vec{L}_{\n}=-i\left(\n\times\vec{\vec{\heartsuit}}_{\n}\right)$ and its square 
$-\vec{\vec{\heartsuit}}^2_{\n}=\vec{L}^2_{\n}\equiv {\cal L}_{\n}$, and the  self-adjoint 
operator $\Lambda_{\n}=\sqrt{{\cal L}_{\n}+\frac 14}-\frac 12$, 
the Green function admits the following operator form of asymptotic expansion \cite{k_Sn,k_t} 
for $|\xv|=\varrho<R$  
\begin{eqnarray}
&&\!\!\!\!\!\!\!\!\!\!\!\!\!\!\!\!\!\!\!\!\!\!\!\! 
\frac{e^{\pm ik|\vec{\rm R}-\xv|}}{4\pi|\vec{\rm R}-\xv|}=
\frac{\chi_{{\Lambda}{\!}_{\n}}(\mp ikR)}{4\pi R}
e^{\mp ik(\n\cdot\xv)}\sim
\frac{e^{\pm ikR}}{4\pi R} \left\{1+\sum^\infty_{S=1}\frac{ \displaystyle 
\prod\limits^S_{\mu=1}\left[{\cal L}_{\n}-\mu(\mu-1)\right]}
{S!(\mp 2ikR)^S}\right\} e^{\mp ik(\n\cdot\xv)}.   
\label{18} 
\end{eqnarray}
Here the eigenvalue ${\cal L}_{\n}\mapsto l(l+1)$, with $\Lambda_{\n}\mapsto l$ for integer 
$l$, and the function $\chi_{l}(y)$ is a sort of well known ``spherical''  Macdonald function 
\cite{T,Ng,blp,g_r,k_Sn,k_t}).  
When the potential\footnote{It is enough for it to have finite first absolute moment and to 
decrease at $\varrho\to\infty$ faster than any power of $1/\varrho$ \cite{k_Sn}.} $U(\varrho)$ 
has a finite effective range $\varrho_0$, this expansion allows us to calculate all the 
pre-asymptotic inverse-power corrections to the wave function 
$\Psi^\pm_\ka(\R)$. Thus, it is possible to write the  differential scattering flux $d\Sigma(R)$ as 
asymptotic series on $R^{-S}$, similar to (\ref{18}), at $R\gg\varrho_0$ \cite{k_Sn,k_t},   
\begin{eqnarray}
&&\!\!\!\!\!\!\!\!\!\!\!\!\!\!\!\!\!\!\!\!\!\!\!\! 
\Psi^\pm_\ka(\R)\, \underset{R\gg\varrho_0}{\sim}\,e^{i(\ka\cdot\R)}+
\frac{{\chi}_{{\Lambda}{\!}_{\n}}(\mp iqR)}{R}\,f^\pm(q\n;\ka)=
e^{i(\ka\cdot\R)}+\frac{e^{\pm iqR}}{R}\left\{
\sum^\infty_{S=0}\frac{h^\pm_S(q\n;\ka)}{(\mp 2iqR)^S}\right\}_{q=k}, 
\label {29} \\
&&\!\!\!\!\!\!\!\!\!\!\!\!\!\!\!\!\!\!\!\!\!\!\!\! 
\mbox{with }\;
h^\pm_S(k\n;\ka)=\frac{{\cal L}_{\n}-S(S-1)}{S} h^\pm_{S-1}(k\n;\ka)=
\frac{1}{S!}\prod\limits^S_{\mu=1}\left[{\cal L}_{\n}-\mu(\mu-1)\right]f^\pm(k\n;\ka) 
\label{30} \\
&&\!\!\!\!\!\!\!\!\!\!\!\!\!\!\!\!\!\!\!\!\!\!\!\! 
\mbox{for amplitude }\;
f^\pm(\q;\ka)=
-\,\frac{1}{4\pi}\int\! d^3{\rm x}\,e^{\mp i(\q\cdot\xv)}U(|\xv|)\Psi^\pm_\ka(\xv), \quad 
\ka=k\kap,\;\;\q=q\n,\;\;q \mapsto k,
\label{30_0} \\
&&\!\!\!\!\!\!\!\!\!\!\!\!\!\!\!\!\!\!\!\!\!\!\!\! 
\mbox{and }\;
\frac{d\Sigma(R)}{d\Omega(\n)}=\frac{1}{2ik}\left[\overset{*}{f}{}^+(q\n;k\kap)
\left(\chi_{\overset{\leftarrow}{\Lambda}{\!}_{\n}}(iqR)
\overset{\leftrightarrow}{\partial}_{\!R} 
\chi_{\overset{\rightarrow}{\Lambda}{\!}_{\n}}(-iqR)\right)f^+(q\n;k\kap)\right]_{q=k},\;\;
\label {S_20} \\
&&\!\!\!\!\!\!\!\!\!\!\!\!\!\!\!\!\!\!\!\!\!\!\!\! 
\mbox{where }\;
\Sigma(R)\equiv\int\frac{d\Sigma(R)}{d\Omega(\n)}\,d\Omega(\n)=\sigma,\;\mbox{ due to }\;
(\chi_l(ikR)\overset{\leftrightarrow}{\partial}_R \chi_l(-ikR) )=2ik.  
\label {S_21} 
\end{eqnarray} 
The upper arrows indicate the directions of action of the operators, e.g. 
$\overset{\leftrightarrow}{\partial}_{\!R}\!=\!\overset{\rightarrow}{\partial}_{\!R}-
\overset{\leftarrow}{\partial}_{\!R}$ 
and 
$f^*$ is the complex conjugate of $f$. 
The important feature of asymptotic power expansions (\ref{18})--(\ref{S_20}) is that they 
still exactly disappear \cite{k_Sn} for total (elastic) flux $\Sigma(R)$ (\ref{S_21}), 
which turned to cross-section $\sigma$ (\ref{3_}) (already not depending on $R$) due to the 
self-adjointness of operators ${\cal L}_{\n},\,\Lambda_{\n}$ on the unit sphere and the value 
(\ref{S_21}) of Wronskian \cite{T,Ng}. 
It may be noted that the influence of the behavior (\ref{29}) at such finite spatial distance 
$R$ onto event rate seems important for explanation of reactor anomaly in 
neutrino flavor oscillations \cite{k_t,n_sh,NN_shk}. 


We will now analyze a short distance correction to such a system due to the existence of a 
minimal measurable length in spacetime. This will be done by analyzing the effects of  
a minimal measurable length on the asymptotic expansion of scattering wave function (\ref{29}) 
and then on the differential scattering flux (\ref{S_20}). Now the scattering wave function 
$\Psi^\pm_{\lv_j}(\R)$ being a positive energy ($E>0$) solution of Schr\"odinger equation 
would satisfy the Lippmann-Schwinger equation deformed by the existence of a minimal length 
in spacetime. 
The existence of a minimal measurable length scale in turn deforms the usual uncertainty 
principle from $\Delta x\Delta p\geqslant {1}/{2}$ to a generalized uncertainty principle: 
$\Delta \widehat{x}_l \Delta \widehat{p}_l\geqslant|(1+3\beta\MVV{\widehat{p}^2_l})|/2$, 
and $\Delta\widehat{x}_l\Delta\widehat{p}_m\geqslant|\beta\MVV{\widehat{p}_l\widehat{p}_m}|
\geqslant 0$ for $l\neq m$, where $\beta$ is a small perturbative dimension parameter of the 
deformation \cite{2z,14,qft1}. Since $\MVV{\widehat{\p}^2}\geqslant\MVV{\widehat{p}^2_l}$ 
holds for averaging over any quantum state, the generalized uncertainty principle in turn 
deforms the Heisenberg algebra, and deformed operators statisfy 
\cite{17,18,51,55,qft1}: 
\begin{equation}
[\widehat{x}_l, \widehat{p}_m]=
i[\delta_{lm}+\beta(\delta_{lm}\widehat{\p}^2+2\widehat{p}_l \widehat{p}_m)]\approx 
i[\delta_{lm}+\beta(\delta_{lm}{\p}^2+2{p}_l {p}_m)],\quad l,m=1,2,3,
\quad \widehat{\p}^2=\widehat{p}_l \widehat{p}_m \delta_{lm}.   
\label{H_d}
\end{equation}
This deformed Heisenberg algebra of deformed operators $\widehat{x}_l,\,\widehat{p}_m$ 
can be related perturbatively for small $\beta$ to the usual Heisenberg algebra 
$[x_l, p_m] = i \delta_{lm}$, with usual representation of $\p=-i\vec{\nabla}_{\xv}$ for  
$p_l p_m \delta_{lm} = \p^2$, as $\widehat{x}_m= x_m$ and 
$\widehat{\p}=\p(1+\beta\p^2)=-i\vec{\nabla}_{\xv}(1-\beta\vec{\nabla}^2_{\xv})$ 
\cite{17,18,51,55,qft1}.  
It may be noted that other sources of such deformations of the Heisenberg algebra have been 
motivated by non-locality \cite{nonl}, doubly special relativity \cite{dsr, dsr1}, deformed 
dispersion relations in the bosonic string theory \cite{stri}, Horava-Lifshitz gravity 
\cite{HoravaPRD, HoravaPRL}, discrete spacetime \cite{Hooft}, models based on string 
field theory \cite{Samuel1}, spacetime foam \cite{Ellis}, spin-network \cite{Gambini}, 
and noncommutative geometry \cite{Carroll}. 
So, we will deform the coordinate representation of the momentum operator in such a 
general way, that the free Hamiltonian (for $\xv\mapsto\R$) is deformed as 
\begin{equation}
H_0=\p^2=-\vec{\nabla}^2_{\R}\longmapsto\widetilde{H}_0=(\widehat{\p})^2=
\p^2+\beta g(\p^2)=-\vec{\nabla}^2_{\R}+\beta g(-\vec{\nabla}^2_{\R}), 
\label{Ham_d}
\end{equation}
where $g(-\vec{\nabla}^2_{\R})$ is a real differentiable function of the 
$-\vec{\nabla}^2_{\R}$, such that $g(0)=0=g^\prime(0)$. Strictly speaking, it describes only 
first term of low energy expansion of some more general 
Hamiltonian\footnote{Similar to first relativistic correction from the expansion 
of $E_p=c\sqrt{\p^2+(Mc)^2}-Mc^2\approx \p^2/(2M)-(\p^2)^2/(8M^3c^2)$.
The respective Lagrangian picture for the case (\ref{Ham_d}) corresponding to (\ref{H_d}) 
is discussed below.}. 
Now the deformation (\ref{H_d}) of the Heisenberg algebra, produces a polynomial   of   
order $g(-\vec{\nabla}^2_{\R})\mapsto 2(-\vec{\nabla}^2_{\R})^N$ with $N=2$ in the Hamiltonian (\ref{Ham_d}). 
However, as shown below, the results obtained here can be easily  generalized to 
any general polynomial function $g(z)$ of $z=-\vec{\nabla}^2_{\R}$, containing any powers 
$z^n$ with $2\leqslant n \leqslant N$ (for arbitrary $N>2$). The dimension of $\beta$ is 
always determined by the lowest value of $n$. 

The deformation of the Lippmann-Schwinger equation can be obtained from the deformation of 
stationary Schr\"odinger equation (\ref{4_}). Now in coordinate representation of momentum 
operator the deformation of  the above free Hamiltonian (\ref{Ham_d}),  can be written as 
\begin{eqnarray}
\left[-\widetilde{H}_0+k^2\right]\Psi^\pm_{\lv_j}(\R)=U(R)\Psi^\pm_{\lv_j}(\R). 
\label{4_0}
\end{eqnarray}
Thus, the wave function for this system satisfies the deformed  Lippmann-Schwinger equation 
\begin{eqnarray}
\!\!\!\!\!\!\!\!\!\!\!\!\!\!\!\!\!\!\! 
\Psi^\pm_{\lv_j}(\R)=e^{i(\lv_j\cdot\R)}-
\int\! d^3{\rm x}\,{\cal G}^{(\pm)}_k(\vec{\rm R}-\xv)\,
U(|\xv|)\Psi^\pm_{\lv_j}(\xv). 
\label{5_0} 
\end{eqnarray}
It depends on solutions of free problem with the assumed entire (polynomial) function 
$g(z)$ as
\begin{eqnarray}
&& 
\widetilde{H}_0\,e^{i(\lv_j\cdot\R)}=k^2\,e^{i(\lv_j\cdot\R)}\equiv
{\cal E}\,e^{i(\lv_j\cdot\R)}, \qquad \lv_j=\ell_j(k)\kap, 
\label{5_1_00} \\
&& 
\left[\vec{\nabla}^2_{\R}-\beta g(-\vec{\nabla}^2_{\R})+k^2\right] 
{\cal G}^{(\pm)}_k(\R-\xv)=-\delta_3(\R-\xv),
\label{5_1} \\
&& 
{\cal G}^{(\pm)}_k(\R-\xv)\equiv 
\int\!\frac{d^3{\rm q}}{(2\pi)^3}\,e^{i(\q\cdot(\R-\xv))}\, F(\q).  
\label{5_2_0} 
\end{eqnarray}
Since Eq. (\ref{5_1}) leads to the equation 
$\left[\q^2+\beta g(\q^2)-k^2\right]F(\q)=1$, the above relation can be expressed as 
\begin{eqnarray}
&&
{\cal G}^{(\pm)}_k(\R-\xv)=\int\!\frac{d^3{\rm q}}{(2\pi)^3}\frac{e^{i(\q\cdot(\R-\xv))}}
{\left[\q^2+\beta g(\q^2)-k^2\mp i0\right]}, \;\mbox { where }\; |\R-\xv|=r.  
\label{5_2} 
\end{eqnarray}
For $\Phi(q^2)=q^2+\beta g(q^2)-k^2$, with $\q=q\n^\prime$ and 
$d^3{\rm q}=q^2dq d\Omega(\n^\prime)$, the Green function is reduced to
\begin{eqnarray}
&& 
{\cal G}^{(+)}_k(\R-\xv)=
\frac{2}{(2\pi)^2 r}\!\int\limits^{\infty}_{0}\frac{dq\,q \,\sin qr}{\Phi(q^2)-i0}=
\frac{1}{i(2\pi)^2 r}\!\int\limits^{\infty}_{-\infty}\!
\frac{dq\,q\,e^{iqr}}{\Phi(q^2)-i0}= 
\nonumber \\
&& 
=\frac{2\pi i}{i(2\pi)^2 r}\sum^N_{s=1}{\rm res}\,
\left(\frac{q\,e^{iqr}}{\Phi(q^2)}\right)\biggr|_{q=\ell_s}=
\frac{1}{4\pi r}\sum^N_{s=1}\frac{e^{i\ell_s r}}{\Phi^\prime(\ell_s^2)},
\label{5_4} \\
&& 
\mbox{where }\; \Phi(q^2)=0,\;\mbox{ for }\; q^2=\ell_s^2, 
\label{5_4_0} \\
&& 
\mbox{with }\; \Phi^\prime(\ell_s^2)=1+\beta g^\prime (\ell^2_s)\neq 0,
\;\mbox{ and }\;Im\,\ell_s>0,\;\mbox{ or }\;\ell_s\mapsto\ell_s+i0, 
\;\mbox{ for }\;Im\,\ell_s=0. 
\label{5_5}
\end{eqnarray}
We only need to consider the first-order poles, because the poles of higher orders will produce positive 
powers of $r$, destroying the asymptotic expansion (\ref{18}). The real polynomial function 
$\Phi(q^2)$ has only real or complex conjugate zeros $\ell^2_s$ (\ref{5_4_0}), which statisfy the 
properties given in (\ref{5_5}). For example, when all the contributions to (\ref{5_4}) (for $s\geqslant 2$)
have $Im\,\ell_s>0$ and vanish exponentially with $r\to\infty$, the only one zero 
(\ref{5_4_0}) $\ell_1=\ell_1(k)>0$ of function $\Phi(q^2)$, admits the radiation condition of 
(\ref{5_5}). So in expansion (\ref{5_4}) for $\xv=\varrho\vs$, 
$r=|\vec{\rm R}-\xv|\gg 1/Im\,\ell_s$ with $s\geqslant 2$, we need to keep only this zero, 
substituting (in the sense of asymptotic expansion)
\begin{eqnarray}
&&\!\!\!\!\!\!\!\!\!\!\!\!\!\!\!\!\!\!\!
{\cal G}^{(+)}_k(\R-\xv)\longmapsto 
\frac{e^{i\ell_1|\vec{\rm R}-\xv|} }{4\pi\Phi^\prime(\ell_1^2)|\vec{\rm R}-\xv|} 
\label{5_50_0} 
\end{eqnarray}
into Eq. (\ref{5_0}).  
Instead of amplitude (\ref{30_0}), it leads to the on-shell scattering amplitude,   
containing the  corrections depending on $\beta$ due to $\ell_1(k)$ and 
$\Phi^\prime(\ell_1^2)$ as  
\begin{eqnarray}
&&\!\!\!\!\!\!\!\!\!\!\!\!\!\!\!\!\!\!\!
f^+_{11}(\q;\lv_1)=-\,\frac{1}{4\pi \Phi^\prime(\ell_1^2)}\int\!d^3{\rm x}\,
e^{-i(\q\cdot\xv)}U(|\xv|)\Psi^+_{\lv_1}(\xv).   
\label{5_50}
\end{eqnarray}  
So, in this situation one can uniquely define the scattering wave function, scattering 
amplitude and differential scattering flux by simple substitutions of 
$\ka\mapsto\lv_1=\ell_1\kap$, i.e., $k\mapsto\ell_1=\ell_1(k)$, $\q=q\n$, $q\mapsto\ell_1$, 
everywhere in Eqs. 
(\ref{18}) -- (\ref{S_21}), with the respective redefinitions of amplitude (\ref{5_50}) and 
incoming flux. We are only interested  in the physical wave function, 
which defines the physical amplitude (and differential scattering flux (\ref{S_20})) 
\begin{eqnarray}
&& \!\!\!\!\!\!\!\!\!\!\!\!\!\!\!\!\!\!\!
\Psi^+_{\lv_1}(\R)\,\underset{R\gg\varrho_0}{\sim}\,  
e^{i(\lv_1\cdot\R)}+
\frac{{\chi}_{{\Lambda}{\!}_{\n}}(-iqR)}{R}\,f^+_{11}(\q;\lv_1)
\underset{R\to\infty}{\longmapsto} 
e^{i(\lv_1\cdot\R)} + \frac{e^{i\ell_1 R}}{R}\,f^+_{11}(\q;\lv_1) +O(R^{-2}). 
\label{1_0}  
\end{eqnarray}
Indeed, there always exists the single perturbative zero of $\Phi(q^2)$ as a solution of Eq. 
(\ref{5_4_0}), which for small $\beta\to 0$ goes to $k^2$ as 
$\ell^2_s=k^2-\beta g(\ell^2_s)\mapsto \ell^2_1\approx k^2-\beta g(k^2)$, while other 
solutions turn to infinity\footnote{That gives an essentially singular point at $\beta=0$ 
for the function $e^{i\ell_s(k)R}$. We call such $\ell_s(k)$ as non-perturbative ones.} like 
$\beta^{-2\epsilon}$ with $\epsilon>0$.  For example,  for the case (\ref{H_d}), (\ref{Ham_d}) 
with $g(z)=2z^2$, $N=2$, $\epsilon=1/2$ and $\xi=\beta k^2$, they are defined with 
$\ell_s(k)=\left[\ell^2_s(k)\right]^{1/2}$ for $s=1,2$, as 
\begin{eqnarray} 
\left.\!\!\!\begin{array}{c} \ell^2_1(k) \\ \ell^2_2(k) 
\end{array}\!\!\right\}=
\frac{-1\pm\sqrt{1+8\xi}}{4\beta}=\frac{2k^2}{1\pm\sqrt{1+8\xi}}=
\left\{\!\!\begin{array}{c} \ell^2_1(k)=k^2(1-2\xi+8\xi^2+...),\;\mbox { for } |\xi|\ll 1, \\ 
\displaystyle \ell^2_2(k)=-\, \frac{1}{2\beta}-\ell^2_1(k)=-\,\frac{k^2}{2\beta\ell^2_1(k)}. 
\;\;\;\qquad \;
\end{array}\!\!\right. 
\label{11} 
\end{eqnarray}
So, for $\beta>0$, $\xi>0$, $\forall\,k^2>0$ with the main branches of all square roots, 
such as $[-\ell^2_s\mp i0]^{1/2}=\mp i\ell_s+0$ (when $\ell_s>0$), we have the above situation 
with $\ell_1(k)>0$, $\ell_2(k)=i|\ell_2(k)|$. 
For $\beta<0$, $\xi<0$, we have $\ell_2(k)>\ell_1(k)>0$ only until $8|\xi|<1$. 
For $8|\xi|>1$, the roots $\ell^2_{1,2}(k)$ are complex conjugate, with $Im\,\ell_{1,2}>0$ 
and there are no scattering waves. 
In general case for $N>2$, the Green function (\ref{5_4}) is also a multivalued function of 
scattering energy ${\cal E}=k^2$ that admits an extraction of a single-valued branch in the 
cutted ${\cal E}$-plane, with the  cuts for every square root 
$\left[-\ell^2_s(k)\right]^{1/2}$ of the wave numbers (\ref{11}). 
The first physical sheet of its Riemann surface is defined by physical cut \cite{T}, that 
goes along ${\cal E}\geqslant 0$, arising due to square root of perturbative wave number 
$\left[-\ell^2_1(k)\right]^{1/2}$. Then all other cuts with $s\geqslant 2$ are on the next 
sheets. For the above case (\ref{H_d}), (\ref{Ham_d}), with $N=2$, $\epsilon=1/2$, the next 
``kinematical'' cut from the point ${\cal E}=-(8\beta)^{-1}$ (for small $\beta$) lies far away 
from the origin on the first sheet (for $\beta>0$). Besides, it is screened by physical cut 
for $\beta<0$. 

Actually, the free states of scattering theory may be determined by any real branch 
(\ref{5_5}) of spectrum of free Hamiltonian (\ref{5_1_00}). So,  we are dealing   with a sort 
of multichannel problem \cite{T}. The point is that every $s$-term in the sum (\ref{5_4}) 
with $\ell_s$  from Eq. (\ref{5_4_0}), being a solution to free Eq. (\ref{4_}) 
(with $U=0$) for $k\mapsto\ell_s$, is a solution to homogeneous Eq. (\ref{5_1}) for $r>0$, 
but only the full sum (\ref{5_4}) gives the solution (\ref{5_2}) to non-homogeneous Eq. 
(\ref{5_1}) for $r\geqslant 0$. 
After substitution of the relations (\ref{5_4}), (\ref{18}) into Eq. (\ref{5_0}), the 
asymptotic expansion of wave function for arbitrary real $j$-mode $\ell_j=\ell_j(k)$, with 
$1\leqslant j,s \leqslant\overline{N}\leqslant N$ and $\R=R\n$, $\lv_j=\ell_j\kap$, 
$\q_s=q_s\n$, $q_s=\ell_s$, can be written as   
\begin{eqnarray}
&& \!\!\!\!\!\!\!\!\!\!\!\!\!\!\!\!\!\!\!
\Psi^+_{\lv_j}(\R)\,\underset{R\gg\varrho_0}{\sim}\,
e^{i(\lv_j\cdot\R)}+\sum^{\overline{N}}_{s=1}
\frac{{\chi}_{{\Lambda}{\!}_{\n}}(-iq_sR)}{R}\,f^+_{sj}(\q_s;\lv_j) 
\underset{R\to\infty}{\longmapsto} 
e^{i(\lv_j\cdot\R)} +\sum^{\overline{N}}_{s=1} 
\frac{e^{iq_sR}}{R}\,f^+_{sj}(\q_s;\lv_j), 
\label{13} \\
&& \!\!\!\!\!\!\!\!\!\!\!\!\!\!\!\!\!\!\!
\mbox{with the scattering amplitudes: }\; f^+_{sj}(\q_s;\lv_j)= 
-\,\frac{1}{4\pi \Phi^\prime(q_s^2)}\int\!d^3{\rm x}\,
e^{-i(\q_s\cdot\xv)}U(|\xv|)\Psi^+_{\lv_j}(\xv). 
\label{13_0}
\end{eqnarray}
Now they look like a multichannel amplitudes \cite{T}, between different channels (modes), if 
$q_s\neq \ell_j$, i.e. $s\neq j$  (where any real $j$-mode scatters into all other possible 
real $s$-modes).

Let us firstly neglect (for small $\beta$) the difference between the exactly conserved 
current and the usually defined diagonal current \cite{T,Ng,blp,k_Sn}, and assume that
(marking this assumption as (!))  
\begin{eqnarray}
&&\!\!\!\!\!\!\!\!\!\!\!\!\!\!\!\!\!
R^2 d\Omega(\n)\left(\n\cdot\jv_{\lv_j,\lv_j}\left[\Psi(\R)\right]\right) 
\overset{(!)}{\longmapsto} 
R^2 d\Omega(\n)
\frac 1{2i}\left(\overset{*}{\Psi}{}^+_{\lv_j}(\R)\overset{\leftrightarrow}{\partial}_R
\Psi^+_{\lv_j}(\R)\right), \; \mbox{ with } \; 
\overset{\leftrightarrow}{\partial}_R=
\left(\n\cdot\overset{\leftrightarrow}{\vec{\nabla}}_R\right).
\label {S_3} 
\end{eqnarray}
Now the sums (\ref{13}) at bilinear form of $\Psi^+_{\lv_j}(\R)$ in    
(\ref{S_3}) lead to the sum of $R$-dependent interference terms proportional to 
$e^{i(\ell_s-\ell_\nu)R}$. 
However, the usually assumed macroscopic averaging over $R$ over detector volume \cite{T} 
recasts these rapidly oscillating exponentials as  
$\MVV{e^{i(\ell_s-\ell_\nu)R}}=\delta_{s\nu}$. 
Thus, repeating all the steps of \cite{k_Sn}, we come to the form of differential scattering 
flux as diagonal sum of the inclusive expressions (similar to (\ref{S_20})) with 
$k\mapsto\ell_j$, $q\mapsto q_s=\ell_s$. Now  using  (\ref{18})--(\ref{30}) in  
(\ref{13}), up to three leading orders in $(\ell_s R)^{-1}$, and under the assumption (!) in 
(\ref{S_3}), we can write   
\begin{eqnarray}
&&\!\!\!\!\!\!\!\!\!\!\!\!\!\!\!\!\!
\frac{d\Sigma_j(R)}{d\Omega(\n)} 
\overset{(!)}{\longmapsto} 
\sum^{\overline{N}}_{s=1} \frac{q_s}{\ell_j}\left\{ 
\biggl|f^+_{sj}(q_s\n;\ell_j\kap)\biggr|^2-
\frac 1{(q_s R)}\,Im\left(\overset{*}{f}{}^+_{sj}(q_s\n;\ell_j\kap)
{\cal L}{}_\n f^+_{sj}(q_s\n;\ell_j\kap)\right)+ \right. 
\nonumber \\
&&\!\!\!\!\!\!\!\!\!\!\!\!\!\!\!\!\!
\left. 
+\frac 1{4(q_s R)^2}\left[\biggl|   
{\cal L}{}_\n f^+_{sj}(q_s\n;\ell_j\kap)\biggr|^2 -
Re\left(\overset{*}{f}{}^+_{sj}(q_s\n;\ell_j\kap)
{\cal L}{}^2_\n f^+_{sj}(q_s\n;\ell_j\kap)\right)\right]+
O\left(\frac 1{(q_s R)^3}\right)\right\}.
\label {S_6} 
\end{eqnarray}
In accordance with (\ref{S_21}), every correction here disappears separately under integration 
over solid angle due to the self adjointness of operator ${\cal L}_\n$ on the unit sphere.  
For $g(z)=2z^N$ one has $1/\epsilon =2(N-1)$. For very small dimensionless deformation 
parameter $\xi=\beta k^{2(N-1)}\ll 1$, every real non-perturbative wave number $\ell_s>0$, 
$s\geqslant 2$ in sum (\ref{13}) turns to infinity like $\ell_s\sim|\beta|^{-\epsilon}$
(similarly, $\ell_2(k)$ (\ref{11}) for $\beta<0$). At the first sight, this leads to extremely 
rapid oscillations of $\exp\{i\ell_s R\}$, that can  be neglected in 
(\ref{13}), leaving us again only with the unique scattering solution (\ref{5_50}), 
(\ref{1_0}). 

However, this is not the case. Up to the values of scattering amplitudes, 
the contributions of these wave numbers in asymptotic expansion (\ref{S_20}), (\ref{S_6}) 
are suppressed again only by the inverse powers of $\ell_s R$. 
Moreover, since the Born approximation becomes relevant for the wave functions (\ref{13}) 
$\Psi^+_{\lv_j}(\R)\approx e^{i(\lv_j\cdot\R)}$ for $j\geqslant 2$ as well as for the 
scattering amplitude (\ref{13_0}), the last arises as real Fourier image of a real function 
$U(\varrho)$, and gives zero contributions to the second term in (\ref{S_6}) of order 
$(\ell_s R)^{-1}$ (as well as of order $(\ell_s R)^{-3}$). 
For the potentials $U(\varrho)$, non-singular at $\varrho\to 0$ \cite{T,Ng}, 
the amplitudes disappear fast enough with the square of momentum transfer 
$\vec{\rm Q}^2_{sj}=(\q_s-\lv_j)^2\to\infty$: 
\begin{eqnarray}
&&\!\!\!\!\!\!\!\!\!\!\!\!\!\!\!\!\!
f^+_{sj}(\q_s;\lv_j)\approx f^{+B}_{sj}(\q_s;\lv_j)= 
-\,\frac{1}{4\pi \Phi^\prime(q_s^2)}\!\int\!d^3{\rm x}\,
e^{- i(\vec{\rm Q}_{sj}\cdot\xv)}U(|\xv|),\;\;\;\vec{\rm Q}_{sj}=\q_s-\lv_j, \;
\mbox{ e.g. for}
\nonumber \\
&&\!\!\!\!\!\!\!\!\!\!\!\!\!\!\!\!\!
\;U(\varrho)=\alpha \varrho^{2\eta-2},\;\;\eta>0,\;
\mbox{ that is }\;  
f^{+B}_{sj}(\q_s;\lv_j)=-\alpha\sin(\pi\eta)\Gamma(2\eta)|\vec{\rm Q}_{sj}|^{-1-2\eta}
[\Phi^\prime(q_s^2)]^{-1}, 
\label{Brn}
\end{eqnarray}  
where for $s\neq j\geqslant 2$ one has 
$|\vec{\rm Q}_{sj}|\sim\ell_j\sim|\beta|^{-\epsilon}\to\infty$. 
Since $\vec{\rm Q}^2_{sj}=\ell^2_s+\ell^2_j-2\ell_s\ell_j\cos\vartheta$ for 
$\kap=\vec{\rm e}_3$, then the operator ${\cal L}_\n$, being for the spherically symmetric 
case (\ref{13_0}) the second order differential operator with respect to the $\cos\vartheta$ 
only \cite{T}, will be the similar operator with respect to the $\vec{\rm Q}^2_{sj}$ 
for every term in (\ref{S_6}).   

To make a self consistent calculations one should take into account the change of conserved 
current supplementing the change in the free Hamiltonian (\ref{Ham_d}). 
Now we consider the  corrections to the currents, for the case with $g(z)=2z^2$,  due to 
difference of exact conserved current from the above assumed simple one (!) (\ref{S_3}). 
For the general field theory, with the higher (second) derivatives the Lagrangian of complex 
classical scalar field depends on its variables as 
${\cal F}={\cal F}(\psi,\psi^*;\partial_\mu\psi,\partial_\mu\psi^*;
\partial_\lambda\partial_\gamma\psi,\partial_\lambda\partial_\gamma\psi^*)$. 
Now for  the field variation   
$\delta\psi$, we can write $\delta(\partial_\mu\psi)=\partial_\mu(\delta\psi)$. So,   the 
variation of the action  
$\delta{\cal I}[\psi,\psi^*]=\delta_\psi{\cal I}+\delta_{\psi^*}{\cal I}$ (for the action 
${\cal I}[\psi,\psi^*]=\int d^4x{\cal F}(\psi,\psi^*;\ldots)$, and using 
$\partial_\mu=(\partial_0,\vec{\nabla}_{\xv})$), can  be expressed as \cite{blp,Uizem}  
\begin{eqnarray}
&&\!\!\!\!\!\!\!\!\!\!\!\!\!\!\!\!\!
\delta{\cal I}[\psi,\psi^*]=\int d^4x\sum_{\varphi\,=\,\psi^*,\psi}
\delta\varphi \left[\frac{\delta{\cal F}}{\delta\varphi}-
\partial_\mu\left(\frac{\delta{\cal F}}{\delta(\partial_\mu\varphi)}\right)+
\partial_\lambda\partial_\gamma\left(\frac{\delta{\cal F}}
{\delta(\partial_\lambda\partial_\gamma\varphi)}\right)\right]+
\nonumber \\
&&\!\!\!\!\!\!\!\!\!\!\!\!\!\!\!\!\!
+\int d^4x\,\partial_\mu\sum_{\varphi\,=\,\psi^*,\psi}\left\{\delta\varphi
\left[\frac{\delta{\cal F}}{\delta(\partial_\mu\varphi)}-
\partial_\gamma\left(\frac{\delta{\cal F}}{\delta(\partial_\gamma\partial_\mu\varphi)}\right)
\right]+\frac{\delta{\cal F}}{\delta(\partial_\mu\partial_\gamma\varphi)}\,
\partial_\gamma(\delta\varphi)\right\}. 
\label {L_2}
\end{eqnarray}
The vanishing of the expressions in square brackets in first integral  in (\ref{L_2}),  
represent the equations of motion. 
The vanishing of the expression for four-divergence in second integral in (\ref{L_2})),  
defines the respective conserved current. Now the equation of motion for the  
Schr\"odinger fields $i\partial_0\psi=(\widetilde{H}_0+U(\xv))\psi$  \cite{blp}, corresponding 
to Schr\"odinger Eq. (\ref{4_0}), with the Hamiltonian (\ref{Ham_d}), for $g(z)=2z^2$, is 
produced by the first square brackets of Eq. (\ref{L_2}), with the following non-relativistic 
Lagrangian, 
for suitable  chosen units and with  $n,m=1,2,3$ 
\begin{eqnarray}
&&\!\!\!\!\!\!\!\!\!\!\!\!\!\!\!\!\!
{\cal F}=i(\psi^*\partial_0\psi-\psi\partial_0\psi^*)-
\left((\vec{\nabla}\psi^*)\cdot(\vec{\nabla}\psi)\right)-U(\xv)\psi^*\psi-
2\beta(\nabla_n\nabla_m\psi^*)(\nabla_n\nabla_m\psi).
\label {L_3} 
\end{eqnarray}
So, for the global gauge transformation \cite{blp}, $\delta\psi=i\psi\delta\alpha$, 
$\delta\psi^*=-i\psi^*\delta\alpha$, $\partial_\mu\delta\alpha=0$, the   
second integral in Eq. (\ref{L_2}) define the gauge current $J^\mu=(J^0,\jv)$ with 
$J^0[\psi]=\psi^*\psi$ and 
\begin{eqnarray}
&&\!\!\!\!\!\!\!\!\!\!\!\!\!\!\!\!\!
2i\jv[\psi]=\left[\psi^*\vec{\nabla}_{\R}\psi-
2\beta\,\psi^*\vec{\nabla}_{\R}(\vec{\nabla}^2_{\R}\psi)+
2\beta\,(\nabla_m\psi^*)\vec{\nabla}_{\R}(\nabla_m\psi)\right]-\left[\ldots\right]^*,
\;\mbox{ that is}  
\label {L_4} \\
&&\!\!\!\!\!\!\!\!\!\!\!\!\!\!\!\!\!
2i\jv[\psi]=\left[\psi^*\vec{\nabla}_{\R}\psi-
4\beta\,\psi^*\vec{\nabla}_{\R}(\vec{\nabla}^2_{\R}\psi)+
2\beta\,\nabla_m\left(\psi^* \vec{\nabla}_{\R}(\nabla_m\psi)\right)\right]-
\left[\ldots\right]^*. 
\label {L_5} 
\end{eqnarray}
This is exactly conserved even for non-diagonal case  because   
$(\vec{\nabla}\cdot\jv_{\lv_j\lv_s}[\psi])=0$ for any stationary scattering solutions 
$\psi=\Psi^\pm_{\lv_j}(\R)$ to Schr\"odinger Eq. (\ref{4_0}). 
It is easy to see that the full three-divergence in the third term in Eq. (\ref{L_5}) does 
not give any contribution to  the incoming flux, 
$\jv[e^{i(\lv_j\cdot\R)}]=\lv_j(1+4\beta\ell^2_j)$. 
Moreover, at least up to the order of $(\ell_s R)^{-4}$, this third term does not give any  
contribution to the differential scattered flux (\ref{S_6}) generated by radial scattered 
flux Eq. (\ref{S_3}). This is because for the current (\ref{L_5}), there is 
$\left(\n\cdot\jv[e^{iq_s R}/R]\right)=q_s(1+4\beta q^2_s)$, and  every summand of 
the sum over $s$ in the first pre-asymptotic relation 
(\ref{13}) (and  every $s$-term in the sum (\ref{5_4})), satisfies \cite {k_Sn} free 
equation (\ref{4_}) as $(\vec{\nabla}^2_{\R}+q^2_s)\psi_{q_s}(\R)=0$ for $\R\neq 0$.  
Eventually, for the case with $g(z)=2z^2$, up to the order of $(\ell_s R)^{-4}$, these 
currents corrections only renormalize the external multipliers of differential scattering 
flux (\ref{S_6}) as  
\begin{equation}
\frac{q_s}{\ell_j}\longmapsto
\frac{q_s}{\ell_j}\frac{(1+4\beta q^2_s)}{(1+4\beta\ell^2_j)}=
\frac{q_s}{\ell_j}\frac{\Phi^\prime(q^2_s)}{\Phi^\prime(\ell^2_j)}\equiv 
\frac{q_s}{\ell_j}\frac{\widetilde{H}^\prime_0(q^2_s)}{\widetilde{H}^\prime_0(\ell^2_j)}=
\frac{v_s}{v_j},\;\mbox { where }\; 
v_j=\frac {\partial\widetilde{H}_0(\ell^2_j)}{\partial\ell_j}.
\label {L_6}
\end{equation}
Here $v_j$ is the velocity. This result looks quite general and conforms to the physical 
meaning of currents, differential scattering flux and cross-section. Now with this 
substitution (denoted below by superscript ${}^{ren}$) we can use the expression (\ref{S_6}) 
without the assumption (!) of (\ref{S_3}). Thus, we can use at least two additional terms of 
orders of $(\ell_s R)^{-3}$ and $(\ell_s R)^{-4}$. They are obtained from the general 
expression (\ref{18})--(\ref{S_20}), with the following replacement in (\ref{S_6})  
\begin{eqnarray}
&&\!\!\!\!\!\!\!\!\!\!\!\!\!\!\!\!\!
O\left(\frac 1{(q_s R)^3}\right)\longmapsto\frac{1}{3(2q_sR)^3}
Im\left[f^*_{sj}{\cal L}{}^3_{\n}f_{sj}-
3\left({\cal L}{}_{\n}f_{sj}\right)^*{\cal L}{}^2_{\n}f_{sj}-
2f^*_{sj}{\cal L}{}^2_{\n}f_{sj}\right]+
\nonumber \\
&&\!\!\!\!\!\!\!\!\!\!\!\!\!\!\!\!\!
+\frac{1}{12(2q_sR)^4}\biggl\{3\bigl|{\cal L}{}^2_\n f_{sj}\bigr|^2+
Re\left[f^*_{sj}{\cal L}{}^4_{\n}f_{sj}-
4\left({\cal L}{}_{\n}f_{sj}\right)^*{\cal L}{}^3_{\n}f_{sj}\right]+ 
12\left[Re\left({f}{}^*_{sj}{\cal L}{}^2_\n f_{sj}\right)-
\bigl|{\cal L}{}_\n f_{sj}\bigr|^2\right]-
\nonumber \\
&&\!\!\!\!\!\!\!\!\!\!\!\!\!\!\!\!\!
-8 Re\left[f^*_{sj}{\cal L}{}^3_{\n}f_{sj}-
\left({\cal L}{}_{\n}f_{sj}\right)^*{\cal L}{}^2_{\n}f_{sj}\right]\!\biggr\}, 
\,\mbox{ where for short: }\, f_{sj}=f^+_{sj}(q_s\n;\ell_j\kap), \;\;
\left({\cal L}{}_{\n}f_{sj}\right)^*=f^*_{sj}\overset{\leftarrow}{\cal L}{}_{\n}.
\label {S_6_1} 
\end{eqnarray}
We see that for non-perturbative modes $\ell_j(k)$ with $j\geqslant 2$, we only have the real 
Born amplitudes of type (\ref{Brn}). They can contribute only to the even powers of $R^{-S}$ 
with $S=0,2,4,\ldots$, in expansion of differential scattering fluxes.   
The respective expressions (\ref{S_6}), (\ref{S_6_1}), renormalized by the 
replacement (\ref{L_6}), are the main result of this work. Rewriting them with obvious 
notations of the summands $d\Sigma^{ren}_{sj}(R)/d\Omega(\n)$ as 
\begin{eqnarray}
\frac{d\Sigma^{ren}_j(R)}{d\Omega(\n)}=
\sum^{\overline{N}}_{s=1}\frac{d\Sigma^{ren}_{sj}(R)}{d\Omega(\n)},\;\,
\mbox { we define also }\;\, 
\overline{\frac{d\Sigma^{ren}(R)}{d\Omega(\n)} }=
\sum^{\overline{N}}_{j=1}\frac{d\Sigma^{ren}_j(R)}{d\Omega(\n)}.  
\label {S_23} 
\end{eqnarray} 
It should be stressed that these quantities actually are those that measured experimentally 
at a finite distance $R$ as differential cross-sections.
Their intrinsic $R$-dependence given here is defined only by observable quantities like the 
on-shell scattering amplitudes (\ref{5_50}), (\ref{13_0}) or partial phase shifts \cite{k_Sn}. 
So, this intrinsic $R$-dependence seems to be sensitive to the 
corrections from existence of the minimal measurable length and can provide an additional 
opportunity for experimental resolution of these corrections.
When the experimental resolution will permit to distinguish between these different 
renormalized differential scattering fluxes (\ref{S_23}) at different $R$, 
the perturbative mode $j=1$ with different $s\geqslant 1$ modes 
(arised in Eqs. (\ref{13}, (\ref{13_0}), (\ref{S_6}), (\ref{S_6_1})) 
will be the most interesting for observation due to the amplification factor (\ref{L_6}). 
Now for the total cross-sections all the inverse-power terms in (\ref{S_6}), (\ref{S_6_1}) 
disappear again, and similar to (\ref{S_21}), we obtain   
\begin{eqnarray}
&&\!\!\!\!\!\!\!\!\!\!\!\!\!\!\!\!\!
\Sigma^{ren}_{sj}(R)=
\frac{v_s}{v_j}\int\left|f^+_{sj}(\ell_s\n;\ell_j\kap)\right|^2d\Omega(\n)={\sigma}^{ren}_{sj}, 
\,\;\mbox{ and }\;\,
\sigma^{ren}_j=\sum^{\overline{N}}_{s=1}{\sigma}^{ren}_{sj}, \qquad 
\overline{\sigma}^{ren}=\sum^{\overline{N}}_{j=1}{\sigma}^{ren}_{j}. 
\label {S_22} 
\end{eqnarray}
So, such a measurement would be interesting when the experimental resolution at least 
between $\sigma^{ren}_{11}$, $\sigma^{ren}_1$, and $\overline{\sigma}^{ren}$ would be achieved.

In this paper, we have analyzed the short distance corrections to scattering processes. 
These corrections occur due to the existence of a minimal measurable length scale in 
spacetime. The existence of a minimal measurable length scale deforms the Heisenberg 
algebra, which in its turn deforms the coordinate representation of the momentum operator. 
The deformation of the momentum operator produces the higher derivative corrections to  
the free Hamiltonian and, besides the changes of different physical processes, 
modifies the Lippmann-Schwinger equation. 
The modification of the Lippmann-Schwinger equation modifies the description of 
scattering processes. We explicitly calculate corresponding corrections to the Green 
function and to conserved current for these processes. 
The obtained modification of scattering amplitudes determine the corrections to the 
observable cross-sections and to the  $R$- dependent differential scattering fluxes defined 
recently in \cite{k_Sn}. 
So, it is justified that the existence of a minimal measurable length regardless to its 
origin can correct scattering processes, and scattering experiments with finite 
macroscopically small base $R$ can in principle detect such corrections. 

It may be noted that the results obtained in this paper are quite general and can be 
applied to most non relativistic scattering processes, where they would act as universal 
corrections to all scattering processes due to an extended structure in spacetime. 
Such corrections would be observed at intermediate scale. 
It will be interesting to use these results to analyze specific scattering processes, and 
to obtain new bounds for the existence of a minimal measurable length scale in spacetime. 

{\bf Acknowledgments} \\
Authors thank M.V. Polyakov and A.E. Rastegin for useful discussions 
and Referee for constructive comments and valuable advice. 


\end{document}